# Polarization Saturation in Multi-layered Interfacial Ferroelectrics

Wei Cao[1#], Swarup Deb[2#], Maayan Vizner Stern[2#], Noam Raab[2], Michael Urbakh[1], Oded Hod[1], Leeor Kronik[3], Moshe Ben Shalom[2]

[1] Department of Physical Chemistry, School of Chemistry, The Raymond and Beverly Sackler Faculty of Exact Sciences and The Sackler Center for Computational Molecular and Materials Science, Tel Aviv University, Tel Aviv 6997801, Israel.

[2] School of Physics and Astronomy, Tel Aviv University, Tel Aviv 6997801, Israel

[3] Department of Molecular Chemistry and Materials Science, Weizmann Institute of Science, Rehovoth 7610001, Israel

[#] These authors contributed equally

**Van der Waals (vdW) polytypes of broken inversion and mirror symmetries were recently shown to exhibit switchable electric polarization even at the ultimate two-layer thin limit. Their out-of-plane polarization was found to accumulate in a ladder-like fashion with each successive layer, offering 2D building blocks for the bottom-up construction of 3D ferroelectrics. Here, we demonstrate experimentally that beyond a critical stack thickness, the accumulated polarization in rhombohedral polytypes of molybdenum disulfide ($r$-MoS$_2$) saturates. The underlying saturation mechanism, deciphered via DFT and self-consistent Poisson-Schrödinger calculations, point to a purely electronic redistribution involving: (1) polarization-induced bandgap closure that allows for cross-stack charge transfer and the emergence of free surface charge; (2) reduction of the polarization saturation value, as well as the critical thickness at which it is obtained, by the presence of free carriers. The resilience of polar layered structures to atomic surface reconstruction, which is essentially unavoidable in polar 3D crystals, potentially allows for the design of new devices with mobile surface charges. Our findings, which are of general nature, should be accounted for when designing switching and/or conductive devices based on ferroelectric layered materials.**

Symmetry manipulation by stacking order control in two-dimensional (2D) crystals has been identified as a route for accessing various ferroelectric phases in 2D material stacks.(*1-11*) For example, broken inversion and mirror symmetries in rhombohedral (r) stacks of van der Waals (vdW) materials made of binary compounds, e.g. boron nitride (BN) and various transition metal dichalcogenides (TMDs), leads to spontaneous interlayer charge transfer and the emergence of interface-confined electric fields. More recently, the same effect, albeit with a smaller magnitude, has been observed in pure graphene stacks.*(12-14)* The interfacial nature of the emerging polarization fields makes the effect robust against external depolarization factors, such as surface effects and doping.(*7*) Such materials, therefore, may serve as 2D building blocks for the bottom-up construction of 3D ferroelectrics.

For few-layered stacks, the interfacial polarization has been found to accumulate with each successive layer by distinct and evenly spaced ladder-like electric potential steps.(*7*) Clearly, such accumulation of polarization cannot proceed indefinitely, as for a thick enough stack the potential drop across the sample will induce cross-stack charge transfer that would inhibit further increase of the polarization.(*15*) Such saturation of the polarization is a general phenomenon that is by no means limited to stacked materials. However, following it experimentally in three-dimensional polar semiconductors, e.g., GaN or AlN, or in oxide ferroelectrics such as $LaAlO_3$, is greatly complicated by the presence of additional effects, e.g., ion mixing, surface and interface reconstruction, the possible presence of contaminants, and the formation of surface defects.(*16-22*) On the contrary, stacked materials comprise of chemically stable individual sheets and therefore offer a unique opportunity for the direct observation of this underlying physical effect. Moreover, by avoiding surface ionic mixing and structural reorganization, the accumulated potential can, in principle, generate mobile surface carriers which were so far limited to epitaxial interfaces of polar semiconductors.

In this study, we follow the evolution of polarization with stack thickness in $MoS_2$ using Kelvin-probe force microscopy (KPFM) measurements and directly observe its saturation. The results are explained via first principles calculations based on density functional theory (DFT). Specifically, the calculations show that the saturated polarization value depends strongly on the presence of free carriers in the 2D stack, which must be accounted for in order to obtain quantitative agreement between theory and experiment.

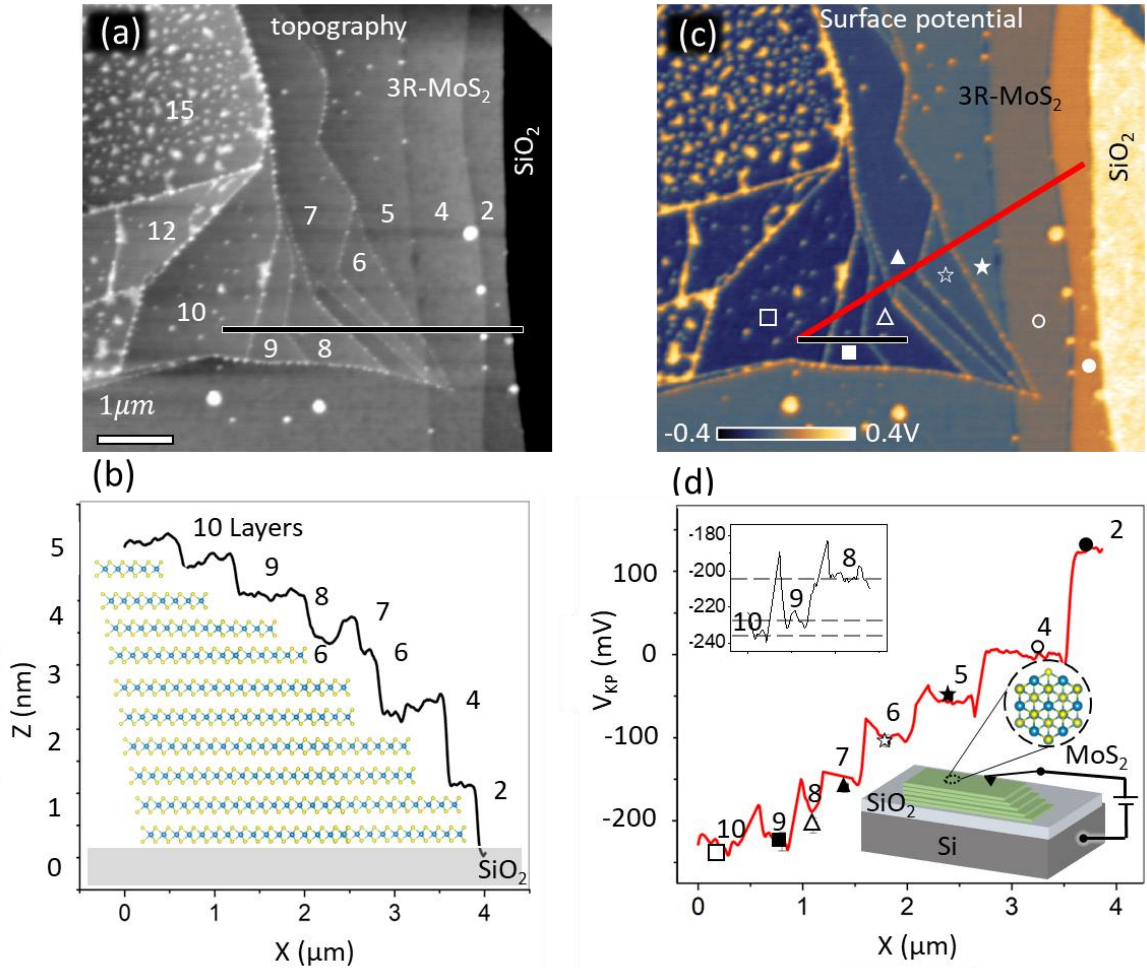

**Fig 1. Surface topography and potential of polar multi-layer MoS$_2$.** (a) AFM topography map of an MoS$_2$ flake exfoliated onto a SiO$_2$ surface. Numbers in white indicate number of MoS$_2$ layers. (b) Topographic cross-section cut along the black line marked in panel a. The number of layers in each step is stated and the stacking sequence is sketched. (c) Surface potential map of the flake shown in panel a, measured by an AFM tip operated at the Kelvin probe mode. (d) Cross sectional cuts of the surface potential map measured along the red (main panel) and black (inset) lines in panel c, with various symbols corresponding to planes designated by the same symbol in panel c. A sketch of the experimental setup is also shown.

To probe the evolution of the spontaneous polarization with the number of layers, we measured the electric surface potential of $r$-MoS$_2$ flakes exfoliated onto a SiO$_2$/Si substrate (see SI sections S1.1-S1.2). An atomic force microscope (AFM) operated at topography mode was used to measure the crystal thickness and deduce the number of layers, $N$ (see Fig.1a, b). The microscope was then used in KPFM mode to scan the same region and obtain a surface potential, $V_{KP}$, map (see Fig. 1c, d, with an additional example given in SI section S1.3). We focus on flakes containing regions of different thickness, ranging from 2-15 layers, all exhibiting co-aligned interfacial polarizations. The crystalline stacking configuration and the measurement setup are illustrated in Fig. 1b, d, respectively. Fig. 1d presents a surface potential line cut that spans stack thicknesses of 2-10 layers. Up to $N = 7$, the potential profile exhibits constant $V_{KP}$ steps of $\Delta V_{KP}$=56 mV per additional layer, indicating a cumulative interfacial effect.(*7*) Notably, above $N = 8$ layers the surface potential steps decrease in size, resulting in an approach towards saturation, as discussed below. The potential jumps appearing near the physical step positions are attributed to surface contamination near step edges, which are manifested in the topography map (Fig. 1a) as bright spots. The surface contamination density increases with stack thickness, possibly due to the increased surface potential (see Fig. 1c). A similar trend has been observed for several other flakes (see results for three more devices in SI section S1.3). To eliminate the effect of surface adsorbates on the reported surface potential, we averaged $V_{KP}$ over clean step surface regions, marked by various symbols in Fig. 1c, where the potential map exhibits a uniform value, ruling out lateral charge distribution effects. The same data are presented also in Fig. 2a (Dev 1), where for convenience the shifted absolute value, $|V_{KP}(N) - V_{KP}(1)|$, is plotted, where $V_{KP}(N)$ is the value obtained over a region of $N$ layers and the potential above a single layer, $V_{KP}(1)$, is extracted as described in SI section S1.4. The potential saturation is found to occur at a value of 0.42 V. Similar measurements of three other devices are further shown (Dev 2,3,4). They exhibit constant potential steps of $\Delta V$ = 78, 63, and 73 meV/layer up to $N \approx 8$, with saturation at 0.57 V, 0.4V , and 0.43 V, respectively. Devices 3 and 4 also show stacking faults with domains of anti-aligned polarizations. (*7, 23*), allowing us to verify the co-aligned nature of the polarization of individual interfaces.

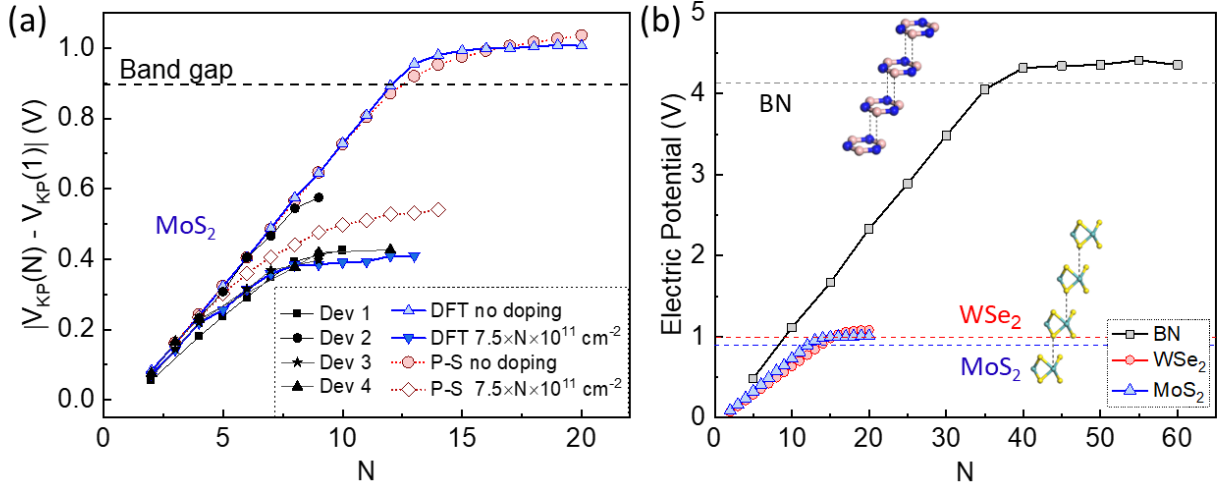

**Fig 2. Surface potential saturation with stack thickness.** (a) Black symbols: measured averaged surface potential changes, $|V_{KP}(N) - V_{KP}(1)|$, as a function of the number of layers, *N*, for four representative *r*-MoS$_2$ flakes, each containing regions of various thicknesses (Dev 1 is the same as in Fig. 1). Blue upright and downward triangles: DFT calculations of the total potential difference across separate *r*-crystals with co-polarized interfaces, under zero doping and under an electronic doping level of $7.5 \cdot 10^{11}$ *e*·cm$^{-2}$ per layer, respectively. The DFT-calculated bulk MoS$_2$ band gap (an underestimate of the experimental one – see text) is given by a dashed horizontal line. Red circle and rhombus symbols: surface potential values obtained via the Poisson-Schrödinger model for the same structures with no doping and under an electron doping of $7.5 \cdot 10^{11}$ *e*·cm$^{-2}$ per layer, respectively. (b) DFT polarization calculations comparing the saturation curves of structurally relaxed *r*-MoS$_2$, *r*-WSe$_2$, and *r*-BN multilayers.

To rationalize these results, we performed DFT calculations of the *r*-MoS$_2$ multilayers (see SI section S2.1), based on the generalized-gradient approximation functional of Perdew, Burke, and Ernzerhof (PBE).(*24*) As shown in SI Section 2.1, we find that atomic relaxation upon stacking is minute, with minor effect on the calculated polarization. This proves that the mechanism underlying polarization saturation is electronic rather than structural. Importantly, the DFT- computed results, shown as blue upright triangles in Fig. 2a, are in qualitative agreement with the experimental results (shown as black symbols for the four devices studied), i.e., showing an initial linear polarization increase with stack thickness followed by polarization saturation. However, quantitative agreement is lacking. The calculated potential saturation value for undoped *r*-MoS$_2$ is roughly twice as large as its experimental counterparts and is obtained at a higher stack thickness of 14 layers, where the potential drop across the stack approaches the PBE-calculated bandgap (~0.9 V) of bulk *r*-MoS$_2$ under periodic boundary conditions. Importantly, the PBE bandgap seriously underestimates the experimental one, ~1.3 eV (*25*), which is a well-known systematic issue.(*26*) It has been previously pointed out (*27-29*) that if the computed bandgap is too small, the critical potential drop at which charge must transfer from the top of the valence band on one surface to the bottom of the conduction band on the other surface, will be reached at a smaller thickness compared to the experimental one. As explained above, it is this charge transfer that causes the saturation of the polarization. Therefore, the computed polarization will saturate prematurely and an underestimated saturated polarization value will follow. Here, however, the exact opposite is found, namely the theoretical values overestimate the experimental ones. Therefore, the discrepancy cannot be attributed to our choice of approximate density functional.

The quantitative discrepancy between experiment and computation can be resolved by considering the effect of free charge carriers in the grown crystals. To examine this, we introduced free electrons in the stacked layers by means of the pseudo-atom approach (see SI section S2.1),(*30*) which we have previously employed successfully in studies of gated stacked layers.(*7*) The results obtained in this case are shown as blue downward triangles in Fig. 2a. Clearly, by introducing a free charge carrier level of $7.5·10^{11}$ $e·cm^{-2}$ per layer, quantitative agreement between theory and experiment (for three of the four devices, with device 2 likely having a lower effective doping level) is obtained. This free carrier level should be considered as a lower bound of the actual carrier level, as it must also compensate for the above-discussed effect of the gap underestimation. To further verify this idea, we performed self-consistent Poisson-Schrödinger (P-S) calculations that simulate the surface potential variation with increasing layer number under extrinsic charge doping (see SI section S3). The results obtained from these calculations, also shown in Fig. 2a, agree well with the DFT calculated polarization curves at both zero doping and at the above-determined free doping level, thereby further confirming our conclusions.

To obtain further insight into the mechanism of polarization saturation, Fig. 3a presents the electronic density of states (DOS) of a 15-layered $r$-MoS$_2$ stack, projected on individual layers. The DOS corresponding to each layer (colored graphs in the figure) is essentially the same as that of the periodic bulk system (gray graph), except for polarization-induced band shifts that, for a thick enough layer, lead to full bandgap closure, i.e., the valence band maximum on the right edge of the stack is at the same energy as the conduction band minimum at its left edge (see further elaboration in SI section S2.2). The same effect is seen by plotting the band structure (Fig. 3b), where each bulk band is expanded into a manifold of bands arising from individual layers, with the VBM and CBM charge densities shown in Fig. 3c exhibiting surface charge characteristics. Once the VBM and CBM achieve the same energy, charge transfer occurs between the two outer layers (see Fig. 3c and SI section S2.3), inducing an electric field that opposes the polarization.(*15*) Upon further increase in the layer thickness, the polarization saturates because the potential drop is pinned by the position of the VBM and CBM. Interestingly, in case of pristine crystals, this charge transfer mechanism beyond a critical thickness is expected to produce mobile electron-like states on one surface and hole-like carriers on the other surface.(*31-33*) A rough estimate for the position of the "knee" is where the shift in DOS owing to the polarization closes the gap. This can be found by equating the overall potential drop prior to saturation, $N \cdot V_I$ (where $N$ is the number of layers and $V_I$ is the potential drop associated with a single interface) with the bandgap, $E_g$. Rounding to the nearest integer, the critical number of stack layers is then $N_c = \text{int}(E_g/V_I) + 1$ (see SI section S2.4 for specific examples). We note that in both experiment and simulations a continuous potential saturation, rather than a sharp "knee", is observed. This is due to the finite electronic temperature and the finite density of states at the band edge.

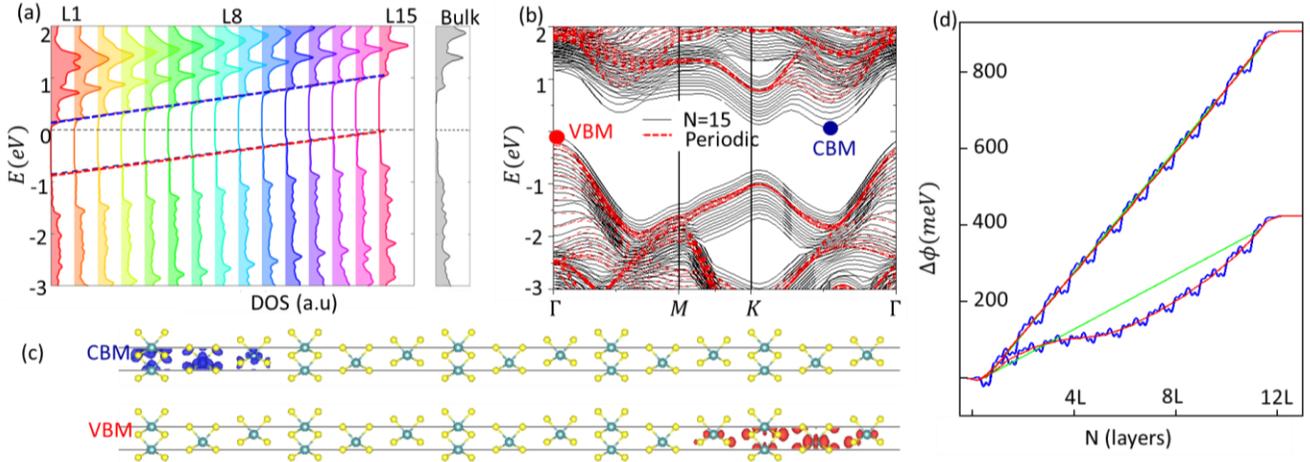

**Figure 3. Demonstration of the electronic mechanism for polarization saturation.** (a) layer-projected DFT-calculated density of states (DOS) for a 15-layered *r*-$MoS_2$ stack (color graphs). The red and blue dashed lines denote the changing position of the local valence band maximum (VBM) and conduction band minimum (CBM) across the stack. The dash-dotted horizontal gray line indicates the Fermi energy. The bulk DOS is given by the gray graph. (b) Electronic band structure of the same system (black lines) compared to its bulk counterpart (red lines). The energy origin is chosen as the VBM, with both VBM and CBM positions marked explicitly. (c) Charge densities corresponding to the VBM and CBM orbitals of the stack, with an iso-surface of $4.8 \cdot 10^{-6}$ e·Å$^{-3}$. (d) Laterally-averaged potential (blue lines) obtained via DFT calculations for an undoped and doped (charge carrier density of $7.5 \cdot 10^{11}$ *e*·cm$^{-2}$) 12-layer *r*-$MoS_2$ stack. The red line denotes the macroscopic potential, obtained from the laterally-averaged one after further perpendicular running-window averaging with a window width of 0.6 nm. The dashed green line demonstrates a linear potential drop. It is virtually indistinguishable from the red line for the undoped case but emphasizes a parabolic- behavior of the red line upon doping. All potentials are presented after subtracting the sum of potential profiles of the corresponding individual monolayers.

The role of doping in the polarization saturation is further clarified by plotting the potential drop across the stack with and without doping, for the case of $N = 12$ – see Fig.3d. The figure shows both the laterally-averaged local potential (blue line) and the macroscopically averaged potential (red line), obtained from the local one by further averaging within a moving window in the direction perpendicular to the stack, with the window width being 0.6 nm, which is the Mo-Mo distance in adjacent layers. Clearly, for the undoped stack the macroscopic potential drop beyond the immediate surface region is essentially linear with position, as expected for an insulator. For the doped case, however, there is a roughly parabolic deviation from the linear drop, which is the hallmark of the emergence of a space charge region with free carriers (*34*). These carriers screen the potential drop, thereby both reducing its final magnitude and promoting the onset of saturation.

The polarization saturation phenomenon, demonstrated here for MoS$_2$ stacks, is of general nature and should be observed for other ladder ferroelectric architectures. Unfortunately, at present 3R crystals of sufficient quality are only available for MoS$_2$, but the effect can still be identified computationally. Fig. 2b compares the DFT calculated polarization curves for undoped *r*-MoS$_2$, *r*-WSe$_2$, and *r*-BN, demonstrating that all three-layered materials exhibit polarization saturation behavior. *r*-BN exhibits a considerably higher saturation polarization (potential drop of ~4.5 V) and a larger saturation thickness, $\sim N = 40$, owing to its much larger bandgap (see SI Sections S2.4-2.5). Similar correlation between gap closure and polarization saturation is found also for 1T'-ReS$_2$ (see SI section S2.6), which was experimentally shown to exhibit ferroelectric behavior.(*35*) These findings validate the mechanism proposed in Ref. (*15*) and agree with similar observations for monolayers of polar oligomers.(*28, 29*) Notably, bandgap closure is not observed for non-polar, anti-parallelly staked AA′ and Bernal AB stacked multilayers, where the gap converges towards its bulk limit with increasing stack thickness (see SI section S2.5). Furthermore, polarization build-up is not observed in semi-metallic stacks, e.g., 1T' WTe$_2$ and MoTe$_2$ (see SI section S2.6), both of which are reported to exhibit interfacial ferroelectricity.(*2, 36*)

In conclusion, we have demonstrated experimentally the emergence of polarization saturation in stacked layers of MoS$_2$. The underlying mechanism, exposed via DFT calculations, agrees with the Ghosez-Gonze-Godby (*15*) suggestion of polarization-induced bandgap reduction with increasing stack thickness, up to a critical point where bandgap closure facilitates charge transfer that suppresses further polarization increase. However, it goes beyond it by showing that doping provides additional screening that reduces the saturated polarization value and stack thickness at which it is attained. This, along with the newly discovered slidetronic switching mechanism, offers new opportunities for the control and design of polarization in novel electronic components.

ASSOCIATED CONTENT

**Supporting Information**. Further details regarding materials and measurements, additional information as to DFT methods and results, and self-consistent solution of the Poisson and Schrödinger equations.

# ACKNOWLEDGMENT

O.H. is grateful for the generous financial support of the Ministry of Science and Technology of Israel (project no. 3–16244) and the Heineman Chair in Physical Chemistry. M.B.S. acknowledges funding by the European Research Council under the European Union's Horizon 2020 research and innovation programme (grant agreement no. 852925), and the Israel Science Foundation under grant nos. 319/22 and 3623/21. O.H. and M.B.S. acknowledge the Centre for Nanoscience and Nanotechnology of Tel Aviv University. L.K. thanks the Aryeh and Mintzi Katzman Professorial Chair and the Helen and Martin Kimmel Award for Innovative Investigation.

# REFRENCES

# Materials and Methods

## S1. Materials and measurements

### S1.1. Device Fabrication

MoS$_2$ flakes of various thicknesses were exfoliated from a bulk 3R crystal, obtained from "HQ Graphene", onto SiO$_2$/Si substrates. Few-nm thick flakes with topographic steps of single layers were selected. An atomic force microscope (AFM) was used to scan the flake's topography.

### S1.2. AFM Measurements

Topography and Kelvin probe force microscopy (KPFM) measurements were performed using a Park System NX10 AFM, employing PPP-EFM n-doped tips with conductive coating. The mechanical resonance frequency of the tip was 75 kHz, its force constant was 3 N/m, and the cantilever was oscillated mechanically with an amplitude of ~20 nm. Amplitude-modulated KPFM (AM-KPFM) measurements are known to produce poor spatial resolution and severely underestimate the measured surface potential, particularly for microscopic samples. Non-local electrostatic interaction of the sample with the cone and cantilever of the scanning tip has been identified as the primary reason behind it. To overcome this issue and to gain a more localized response, we have used frequency-modulated KPFM (FM-KPFM) or sideband KPFM, which is sensitive to electrostatic interaction variation rather than the interaction strength itself.(*1*) Therefore, the sideband signal has the highest contribution from the tip apex, which has the smallest dimension and is the closest to the surface. We note that this type of global external measurement still slightly averages the desired signal with additional signal from the surroundings and thus provides a lower limit to the measured *ΔV* value. In our measurements, we excite the cantilever with an AC voltage amplitude of 2V and a frequency of 2 kHz. The topography and the KPFM signals were obtained separately using a two-pass measurement. The first pass recorded the topography in non-contact mode. In the second pass, the KPFM potential was recorded after lifting the tip an extra 5 nm and following the same topography line-scan, ensuring separation of the topography and the electrical signals. The images in Fig. 1 and S1 were acquired using Park SmartScan software and the data analysis was performed with the Gwyddion program.

## S1.3. Additional Examples

Device 2

Fig. S1 presents an additional demonstration of the polarization saturation for another device (device 2). The noise is analyzed as the standard deviation of the potential at a clean region of a few µm² and is typically less than 20mV (see the error bars in Fig. 1d and Fig. S1). We note that surface contaminants and the noise level tend to accumulate at high potentials and are challenging to avoid, although the inert atmosphere in our microscope.

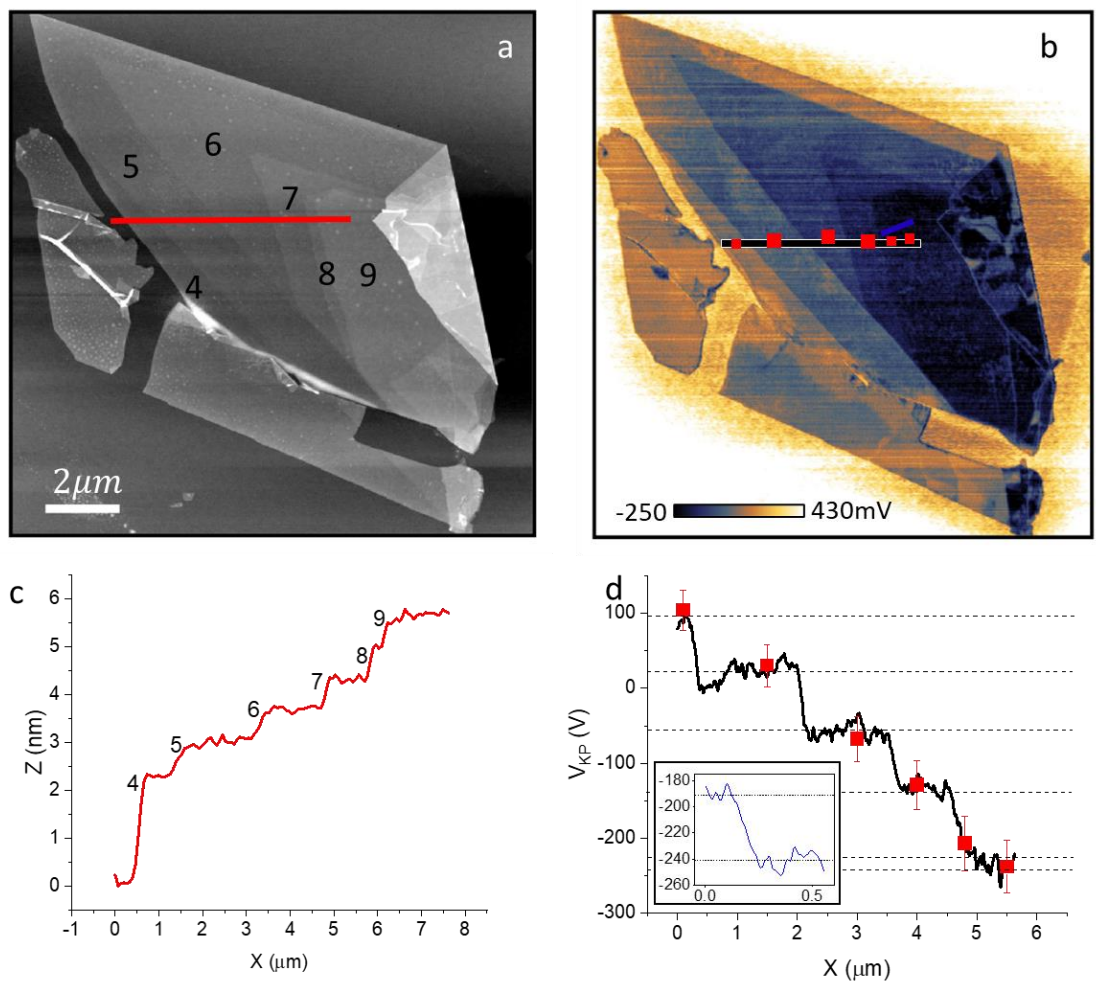

**Figure S1. Additional example of polarization saturation.** Topography (a) and Surface potential (b) maps, measured using AFM and KPFM, respectively. (c) A cross section taken along the red line in (a). (d) A cross section taken along the red (inset: blue) line in b.

Devices 3+4

The flakes corresponding to devices 3 and 4 were transferred onto a graphite substrate and cleaned *in situ* before the KPFM measurements in our vacuum AFM. Fig. S2 presents additional demonstrations of the polarization saturation for devices with non-uniform domains. We observed similar trends in the voltage dependence on the number of layers. We have included the data from these additional devices in Fig. S2 of the supplementary information and incorporated it into Fig. 2a of the main text.

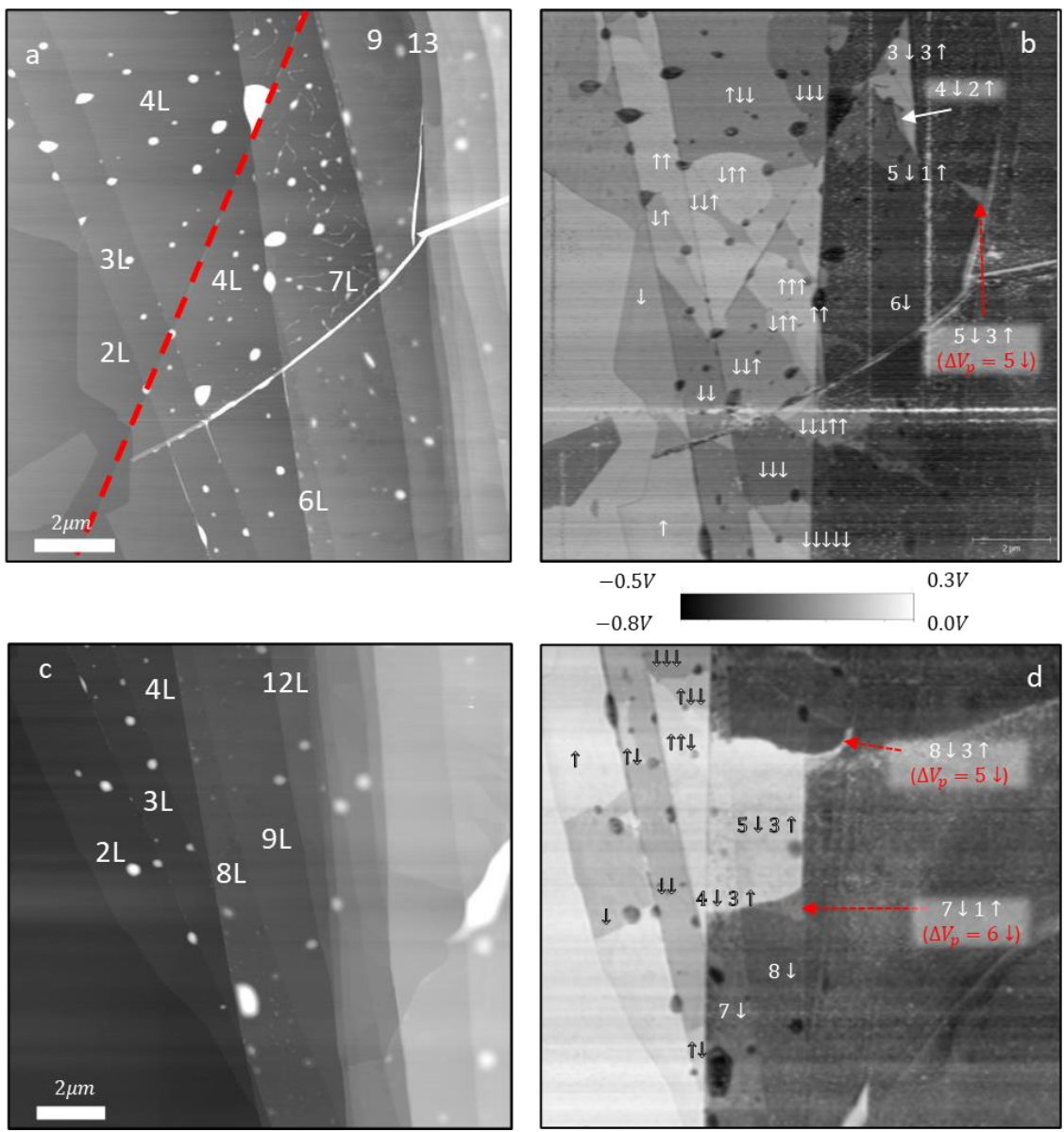

**Figure S2. Additional examples of polarization saturation in multi-domain flakes.** Topography (a,c) and Surface potential (b,d) maps, obtained from AFM and KPFM measurements, respectively, for devices 3 and 4. Here the flakes are placed on a ~ 10 nm graphite flake. The number of layers is annotated on the topography images. In the surface potential maps, arrows indicate the number of polarized

interfaces pointing up and down ($\uparrow, \downarrow$). Note that even for $N > N_{saturation} \approx 8$, as long as $|N_\uparrow - N_\downarrow| < N_{saturation}$, the surface potential accumulates in a linear fashion (see red arrows in b,d).

## S1.4. Comparison of the calculated electric potential with KPFM measurements

To facilitate comparison between the data and the calculated electric potential, the data in Fig. 2a are plotted with shifted absolute values. The calculated electric potential for a stack of $N$ MoS$_2$ layers, $V(N)$, is referenced to $V(1)$. In practice, we shifted all the data relative to the minimum layer measurement, corresponding to $N=2$ for devices 1, 3, 4 and $N=4$ for device 2, with 1 and 3 interfaces, respectively. We then added a correction, $\Delta V$, for changing the reference to $V(1)$. In other words, the plotted data show $|V(N) - V(2)| + \Delta V$ for devices 1, 3, 4 and $|V(N) - V(4)| + 3\Delta V$ for device 2 (where $\Delta V$ is the average measured potential step in the linear regime, the values of which for the four devices are 56, 78, 63, and 73 meV, as also mentioned in the main text).

## S2. Additional details regarding the DFT calculations
### S2.1. Computational details

DFT methods

To obtain the electrostatic potential of the multilayer $r$-MoS$_2$ stack shown in Fig. 2, we used the Perdew-Burke-Ernzerhof (PBE) generalized-gradient exchange-correlation density functional approximation,(*2*) augmented by the Grimme-D3 dispersion correction using Becke-Johnson (BJ) damping,(*3*) as implemented in the Vienna Ab-initio Simulation Package (VASP)(*4*). A plane wave energy cutoff of 600 eV and a k-point mesh of 20×20×1 were used, with a vertical vacuum size of 10 nm to avoid interactions between adjacent images. The core electrons of the atoms were treated via the projector augmented wave (PAW) approach. The 3R stacked structures were constructed and relaxed using the conjugated gradients algorithm with a force threshold of $10^{-3}$ eV/Å. Single-point electron density calculations were then performed on the relaxed structure using a Gaussian smearing of 0.05 eV, to enhance the convergence of the self-consistent cycle. The same approach was used to calculate the potential differences of WSe$_2$ and $r$-BN that are shown in Fig. 2b.

Consistency and convergence tests

To evaluate the vertical polarization, a dipole moment correction was employed.(*5*) For validation purposes, doubled supercell calculations were also performed, yielding nearly identical potential differences (see Fig. S3a). The doubled supercell consists of two opposing mirror images of each MoS$_2$ stack with a 6 nm inter-image vacuum region. Fig. S3b and S3c show the potential profiles along the normal direction calculated by the dipole correction and double supercell methods for N=14, respectively.

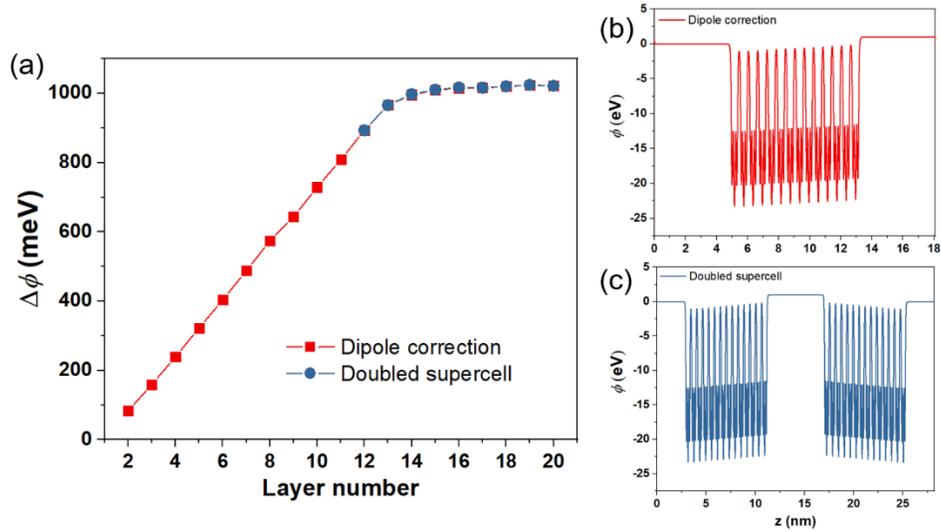

**Figure S3. Potential difference calculations.** (a) Potential drop as a function of number of layers calculated by the dipole correction and a doubled supercell setup for multilayer $r$-MoS$_2$. Potential profiles of the $N$=14 layer system calculated by the (b) dipole correction method and (c) doubled supercell approach.

Convergence tests of the calculations (Fig. S4) indicate that our choice of parameters leads to electrostatic potential differences convergence to within 2.5, 0.1, and 0.7 meV with respect to the number of k-points, energy cut-off, and vacuum size, respectively. The corresponding total energy convergence values are to within 0.04, 8, and 0.7 meV.

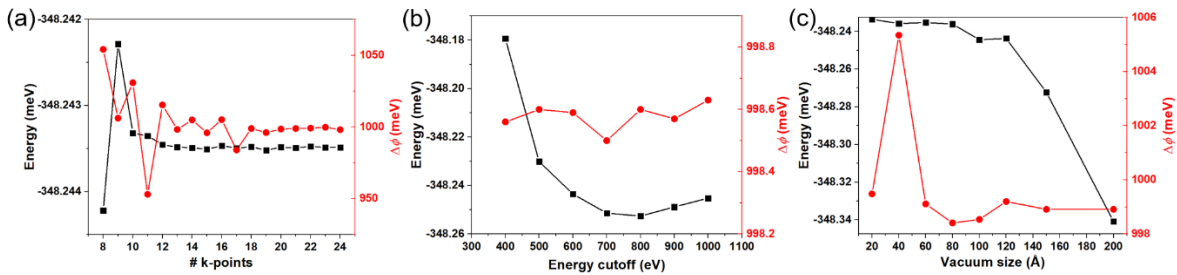

**Figure S4. Convergence tests.** Convergence tests of the total energy (black curve, left vertical axis) and electrostatic potential difference (red curve, right vertical axis) of $N$=16 $MoS_2$ as a function of: (a) number of k-points, (b) energy cutoff, and (c) vacuum size.

Effect of geometry relaxation

Fig. S5 shows a comparison of the potential drop and bandgap calculations for unrelaxed and relaxed multilayer $r$-$MoS_2$. In the former, only the geometry of individual layers is relaxed, as described above, whereas in the latter an additional geometry relaxation step is undertaken for the entire stack. Both the saturation curve and the bandgap dependence on number of layers are weakly affected by stack geometry relaxation, indicating that the polarization saturation mechanism is dominated by electronic, rather than structural, effects.

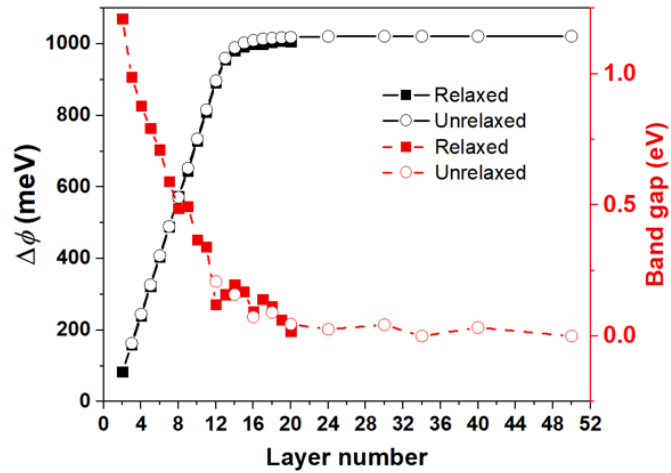

**Figure S5. Effect of geometry relaxation.** The potential difference (black, left) and band gap (red, right) of $r$-$MoS_2$ as a function of layer number before (open symbols) and after (full symbols) stack geometry relaxation.

Hybrid functional and spin-orbit coupling calculations

The polarization saturation shown in the main text was performed at the PBE level of theory, excluding spin-orbit coupling (SOC) effects. To evaluate both the effect of the exchange-correlation density functional approximation and the possible role of SOC contributions, we performed additional single-point polarization calculation using the PBE optimized multilayer $r$-$MoS_2$ model systems, either at the Heyd-Scuseria-Ernzerhof (HSE)(*6-9*) screened-exchange hybrid density functional level of theory, or with scalar-relativistic corrections. Fig. S6 demonstrates that while the HSE bandgap is consistently larger than the PBE one and SOC induces band splitting, at the linear commulative region replacement of PBE by HSE or the inclusion of SOC has an insignificant effect on the calculated polarization.

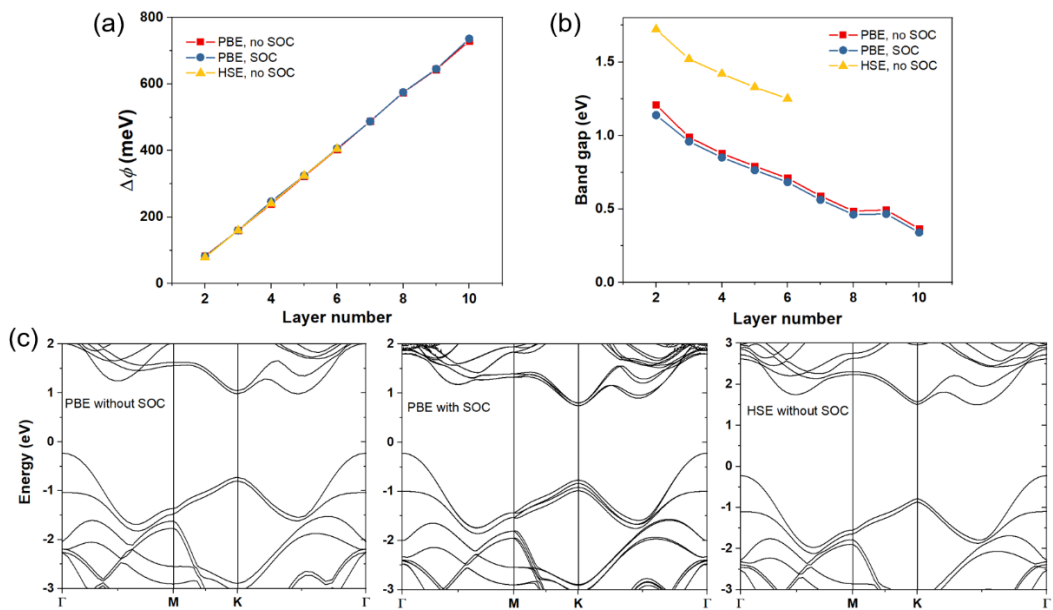

**Figure S6. Screened-hybrid functional and spin-orbit coupling calculations.** (a) Potential drop and (b) bandgap as a function of number of layers obtained using the PBE functional approximation, without (red) and with SOC (blue), and with the HSE functional without SOC (yellow). (c) Corresponding band structures of four-layered $r$-MoS$_2$.

Doping calculations

To perform DFT calculations that include effective doping, multilayer $r$-MoS$_2$ stacks were first geometrically relaxed as described above. This was followed by single-point calculations using the fractional nuclear charge pseudoatom approach,(*10*) as implemented in the Quantum Espresso open source package(*11*). A doping density of $7.5 \times 10^{11}$ cm$^{-2}$ per layer was imposed through fractional charging of both the Mo and S nuclei with an extra charge of $2.16 \times 10^{-4}$ |e|. An energy cutoff of 60 Ry (816.34 eV) was used with a larger k-point grid of $30 \times 30 \times 1$ (found necessary for convergence) and a vertical vacuum size of 10 nm to avoid spurious interactions between adjacent bilayer images. Fermi-Dirac smearing with an effective temperature of ~300 K was used to enhance the convergence of the self-consistent cycle.

To confirm that the calculations employing effective doping do not affect the band structure, we conducted a comparative analysis of the band structure and density of states (DOS) between the undoped and doped 12-layer *r*-MoS$_2$ stack, shown in Fig. S7a. They are additionally compared to the same analysis for non-polar 2H stacked MoS$_2$, as shown in Fig. S7b. In the latter case, minimal band changes are observed, other than a trivial uniform shift, thereby validating the effective doping approach. For the polar layers, doping results in an increase in the band gap, but the DOS curves clearly establish that this is attributed to polarization-induced band shifts, rather than deformation of the bands themselves.

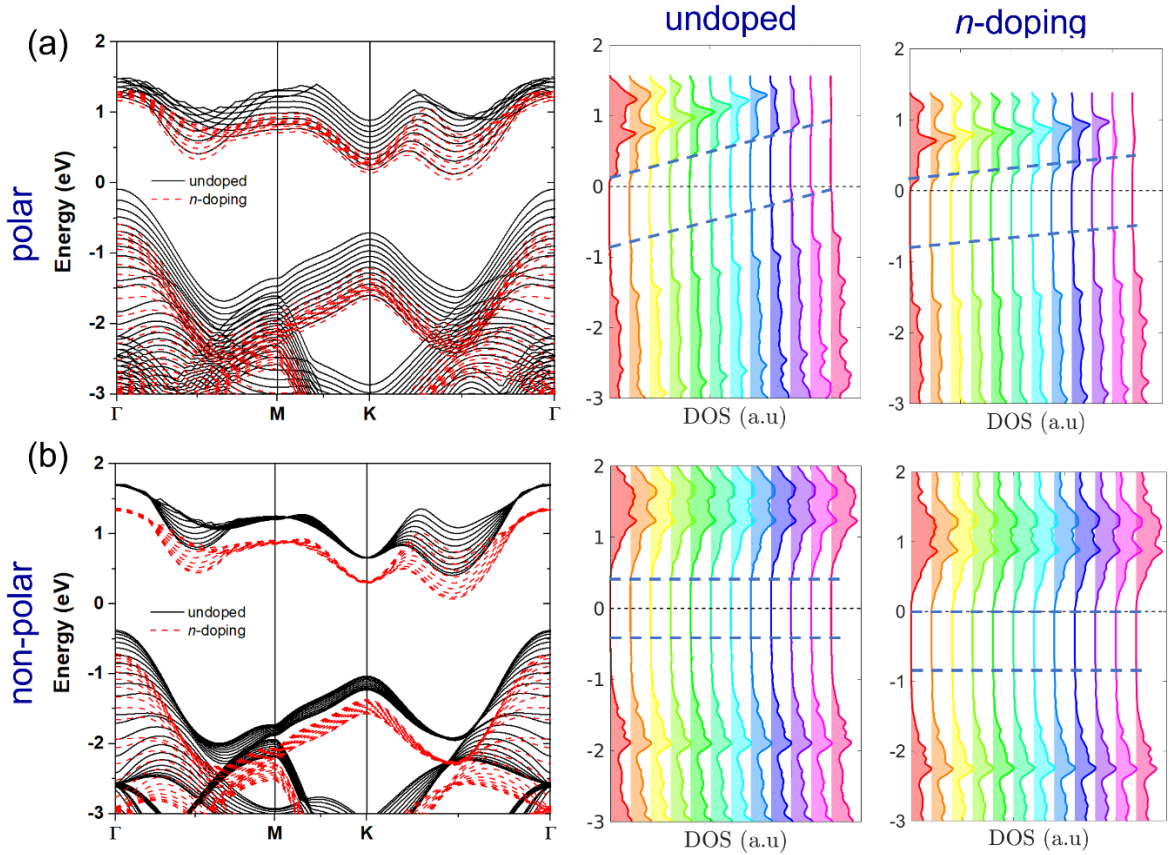

**Figure S7. Demonstration of the validity of doping calculations.** The band structures and density of states (DOS) obtained via DFT calculations for an undoped and doped (charge carrier density of $7.5 \cdot 10^{11}$ $e \cdot cm^{-2}$ per layer) 12-layer (a) rhombohedral (polar) and (b) AA' (non-polar) MoS$_2$ stack. The energy origin is set to the Fermi level. The blue dashed lines denote the changing position of the local valence band maximum (VBM) and the conduction band minimum (CBM) across the stack.

## S2.2. DFT calculated MoS₂ density of states

In Fig. 3a of the main text, we show the DFT calculated density of states projected onto the various layer of a 15-layered $r$-MoS$_2$ stack. For completeness, we present in Fig. S8 similar results for systems of different thickness ($N = 9$, 12, and 18). The slopes of the VBM and CBM spatial variations for the $N = 9$ and 12 systems are similar, whereas for the $N = 18$ system, which is already above the polarization saturation thickness, a lower slope is obtained.

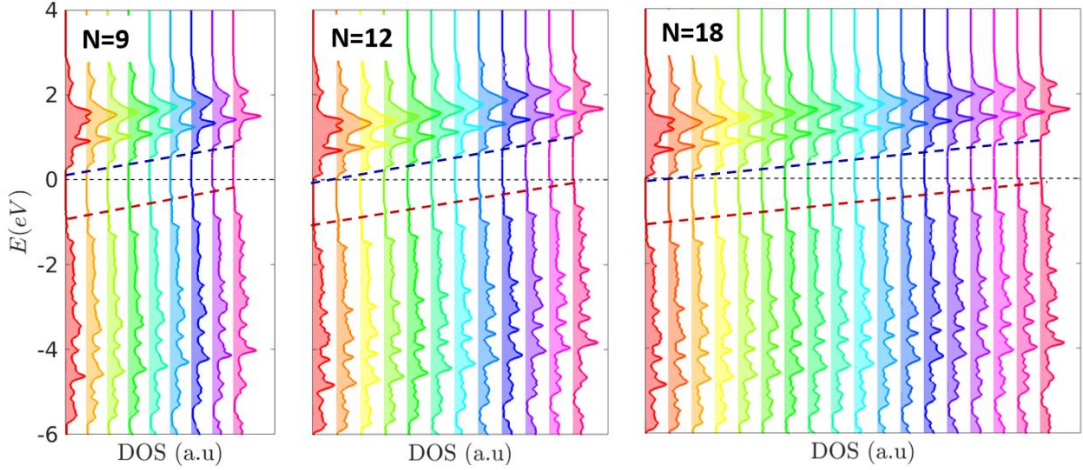

**Figure S8. DFT calculated DOS.** DFT calculated density of states projected onto the various layer of a $r$-MoS$_2$ stack consisting of $N = 9$, 12, or 18 layers.

## S2.3. Potential profile of MoS₂

The charge transfer between the stacked layers is understood by the potential distribution, as illustrated in Fig. S9. The laterally-averaged electrostatic potential profiles, relative to individual monolayers, for $N=12$ (Fig. S9a) and $N=16$ (Fig. S9b) are depicted. Both systems demonstrate uniformly spaced and decoupled potential steps. In the case of the 12-layer configuration, the potential step size matches the potential drop observed in bilayer AB stacked MoS$_2$ (i.e., 83 meV). In contrast, the potential step size in the 16-layer system is smaller, as saturation has already set in.

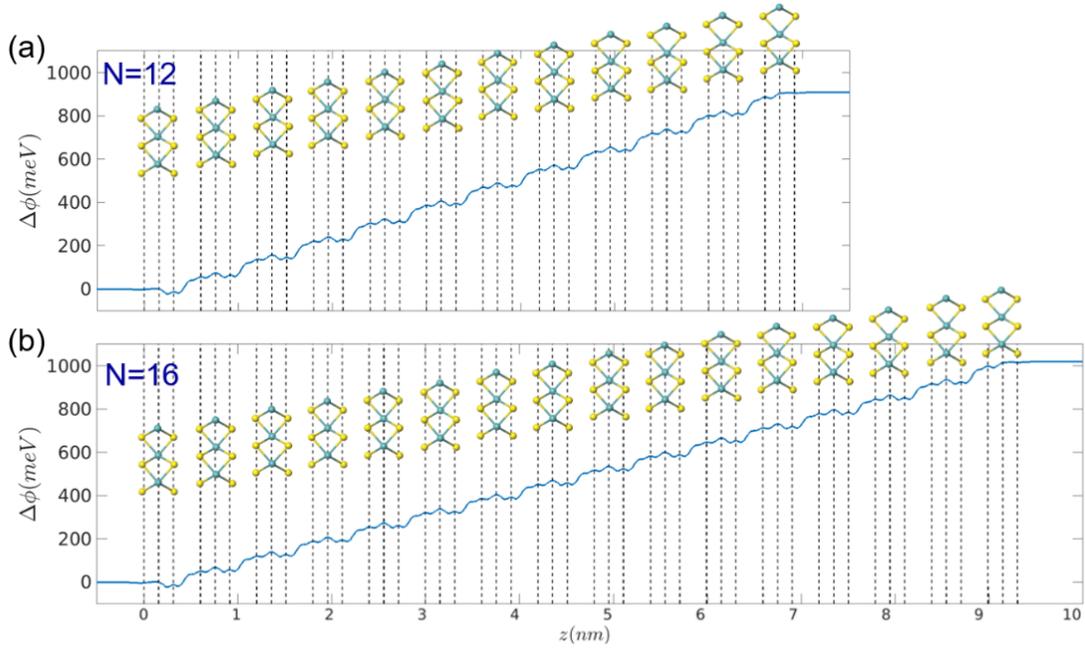

**Figure S9. Potential distributions analysis.** Difference between the laterally-averaged potential profile obtained for $r$-MoS$_2$ with (a) $N = 12$ and (b) $N=16$ layers and that of the corresponding monolayers. The black dashed lines represent the vertical locations of the ions. The origin of the horizontal axis is set to the bottom S sublayer.

### S2.4. Estimation of polarization saturation

Table S1 shows the prediction of polarization saturation for $r$-MoS$_2$, $r$-WSe$_2$, and $r$-BN stacks, yielding the critical thickness, $N_c$, compared to the one predicted from the bandgap.

**Table S1.** Prediction of polarization saturation thickness by the potential drop associated with a single interface, $V_I$, and the bandgap, $E_g$, of a bulk $r$-MoS$_2$, $r$-WSe$_2$, and $r$-BN, under periodic boundary conditions.

| Type | $V_I$ (meV) | $E_g$ (eV) | $N_c$ | |
|---|---|---|---|---|
| | | | Prediction | Calculation |
| MoS$_2$ | 82 | 0.896 | 12 | 14 |
| WSe$_2$ | 69 | 0.992 | 16 | 18 |
| $h$-BN | 115 | 4.136 | 37 | 40 |

## S2.5. Band gap of MoS$_2$, *r*-WSe$_2$ and *r*-BN as a function of stack thickness

In Fig. 2b in the main text we compare the saturation characteristics of *r*-MoS$_2$, *r*-WSe$_2$ and *r*-BN. In Fig. S10a, we present the stack thickness dependence of the bandgap of bernal (labled as AB(A), blue), rhombohedral (labeled as AB(C), yellow), and AA' (red) stacked MoS$_2$. By comparing the polar stacking behavior (yellow) to the non-polar configurations (blue and red), we observe that the latter converge to a finite value, whereas the former approaches bandgap closure following the mechanism discussed in the main text. Similar bandgap closure is seen also for *r*-WSe$_2$ (Fig. S10b) and *r*-BN (Fig. S10c) polar multilayers.

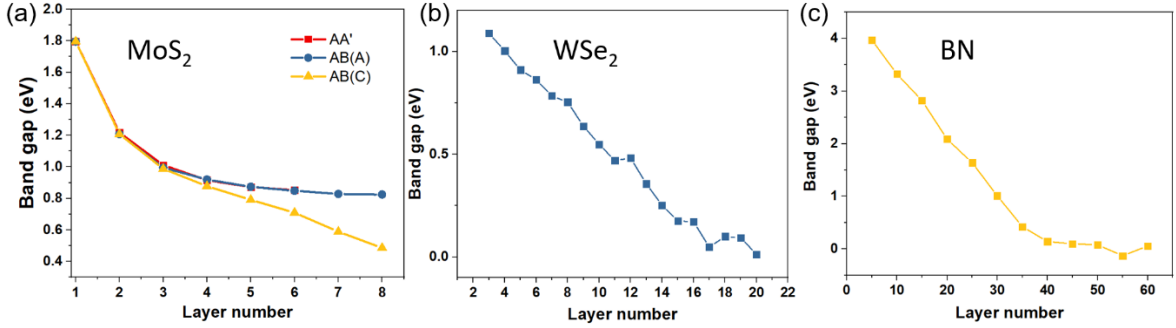

**Figure S10. Band gap of *r*-MoS$_2$, *r*-WSe$_2$, and *r*-BN.** (a) Bandgap of multilayer r-MoS$_2$ at various stacking modes, as a function of number of layers. (b), (c) Bandgap of multilayer *r*-WSe$_2$ and *r*-BN.

## S2.6. Potential drop and bandgap dependence on stack thickness for 1T' ReS$_2$, 1T' WTe$_2$ and MoTe$_2$

The polarization saturation mechanism discussed in the main text is not limited to *r*-MoS$_2$ layered stacks and can be found in other non-centrosymmetrically stacked 2D materials. For example, in 1T'-ReS$_2$, the polarization of which was experimentally studied,(*12*) polarization saturation and bandgap closure are predicted to occur at $N = 18$ (see Fig. S10a). While semi-metallic systems, such as 1T' WTe$_2$ and MoTe$_2$ (Fig. S11b), may exhibit interfacial ferroelectricity,(*13, 14*) no significant dependence on stack thickness is predicted, as expected for zero-bandgap materials.

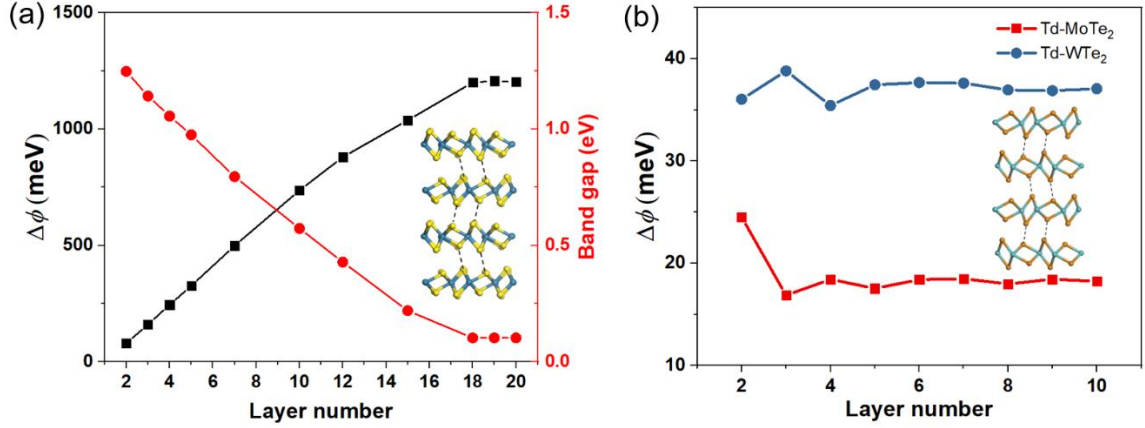

**Figure S11. Potential drop and bandgap as a function of the number of layers for 1T' ReS$_2$, 1T' WTe$_2$ and MoTe$_2$.** (a) DFT-computed potential drop and bandgap calculations of structurally relaxed 1T' ReS$_2$ as a function of stack thickness. (b) DFT-computed potential drop as a function of layer number in semi-metallic 1T' WTe$_2$ and MoTe$_2$ multilayers.

### S3. Self-consistent solution of the Poisson and Schrödinger equations

We used the following algorithm to obtain self-consistent solutions of the Poisson and Schrödinger equations:

1. We start by creating a simplified periodic function to model polarization-induced charge transfer, $\rho_{pol}(z)$. The corresponding polarization-induced potential, $\phi_{pol}(z)$, is then calculated by solving the one-dimensional (1D) Poisson equation:

$$\frac{d^2\phi_{pol}(z)}{dz^2} = \frac{\rho_{pol}(z)}{\epsilon_0}, \quad (1)$$

where $\epsilon_0$ is the free space permittivity. The magnitude and distribution width of $\rho_{pol}(z)$ are then optimized to mimic the DFT-calculated polarization induced potential variation across the stack. An example of the simplified density profile and the corresponding potential profile, compared to the DFT counterparts, is presented in Figs. S12a,b.

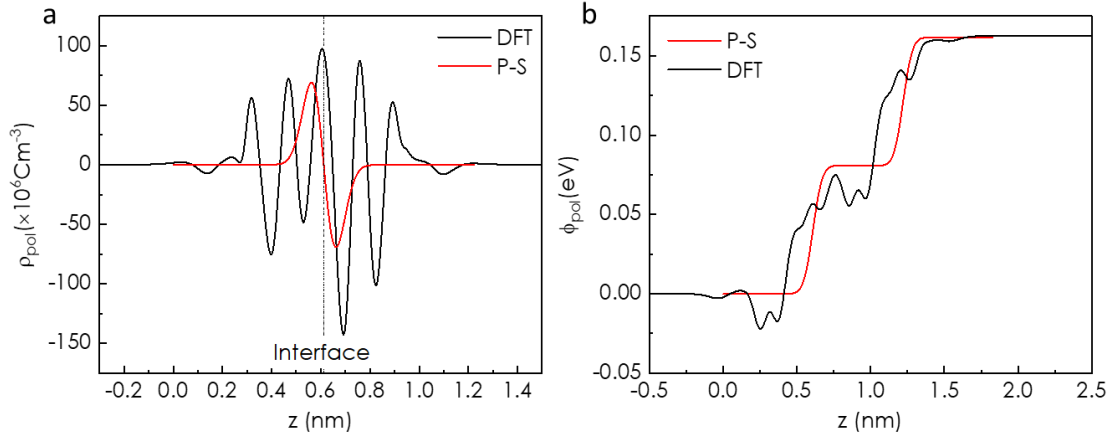

**Figure S12.** (a) Example of a simplified density profile used in the Poisson-Schrödinger (P-S) calculation (red) compared to the DFT result (black). (b) Corresponding potential profiles obtained from the solution of the Poisson equation.

2. A pair of Schrödinger equations are solved to acquire the spontaneously excited free electron and hole densities:

$$-\frac{\hbar^2}{2m_e^*}\frac{\partial \psi_e^2(z)}{\partial z^2} + V_c(z)\psi_e^n(z) = E_e^n \psi_e^n(z) \quad (2)$$

$$-\frac{\hbar^2}{2m_h^*}\frac{\partial \psi_h^2(z)}{\partial z^2} + V_v(z)\psi_h^n(z) = E_h^n \psi_h^n(z) \quad (3)$$

where the conduction and valence band potentials are taken to be $V_c(z) = -e\phi_{pol}(z)$ and $V_v(z) = V_c(z) - E_g$, where $E_g$ is the bandgap of the system, and $E_e^n$ and $E_h^n$ are the n$^{th}$ energy eigenvalues of electrons and holes, respectively. $m_e^*, m_h^*$ are electron and hole effective masses, respectively. We use the effective out-of-plane electron mass at the conduction band $Q$, $0.49 m_e$, and the effective out-of-plane hole mass at the valence band $\Gamma$ point, $0.80 m_e$, (16) where $m_e$ is the free electron rest mass. The free electron and hole densities are then calculated via:

$$\rho_e(z) = -e\, g_v \frac{m_e^* k_B T}{\pi \hbar^2} \sum_n \ln\left(1 + e^{\frac{E_f - E_e^n}{k_B T}}\right) |\psi_e^n(z)|^2, \quad (4)$$

$$\rho_h(z) = e\, g_v \frac{m_h^* k_B T}{\pi \hbar^2} \sum_n \ln\left(1 + e^{\frac{E_h^n - E_f}{k_B T}}\right) |\psi_h^n(z)|^2, \quad (5)$$

where $k_B$ is the Boltzmann constant, T (=300K for our calculation) is the temperature, $\hbar$ is the reduced Planck constant, and $E_f$ is the chemical potential. Eqs. (4) and (5) are obtained via integration of the 2D density of states, multiplied by the Fermi-Dirac distribution, along the two unconfined directions.(15)

The prefactors in the summations, prior to the probability density, accounts for the thermal distribution of free carriers at a finite temperature among available subbands. $g_v$ stands for the valley degeneracy viz. 6 for $Q$ valley of conduction band and 1 for $\Gamma$ valley of valence band.

3. To obtain a self-consistent solution, $\rho_e(z)$ and $\rho_h(z)$ are added to $\rho_{pol}(z)$ in the Poisson equation as follows:

$$\frac{d^2\phi}{dz^2} = \frac{\rho_{pol}(z)}{\epsilon_0} + \frac{\rho_e(z)+\rho_h(z)+e\,N_d}{\epsilon_0\epsilon_r}, \tag{6}$$

where $N_d$ is the density of positively charged donors and $\epsilon_r = 8.9$ is the relative permittivity of bulk MoS$_2$.(*17*)

The process is repeated until self-consistency is achieved. If charge neutrality is satisfied the calculation is terminated, otherwise the value of $E_f$ is modified accordingly and the process restarts. SciPy and NumPy Python-libraries were used to perform these numerical calculations. $E_g$ was used as a free parameter to match the V$_{kp}$ saturation value obtained from DFT. A value of 0.64 eV was used for all the results shown in the paper. It remains a constant for all thickness and doping.

We can further define the bandgap as the energy separation between the frontier eigenvalues in the conduction and valence band (see Fig. S13a). Essentially, it is the correction to the classical bandgap, $E_g$, due to quantum confinement. Interestingly, the bandgap variation as a function of layer number, both for undoped and doped cases, follows a similar trend as that obtained from the DFT calculations.

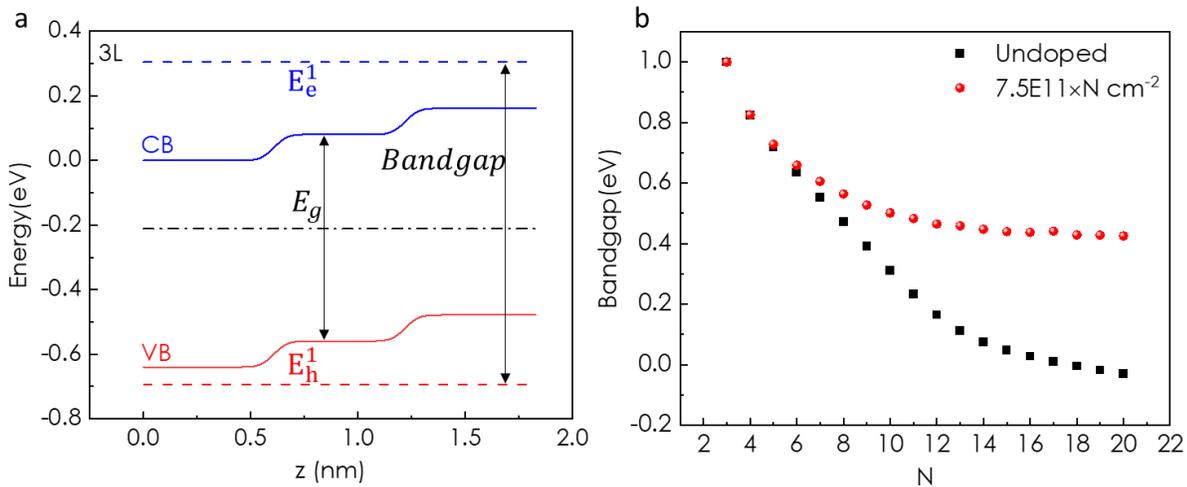

**Figure S13.** (a) Calculated conduction and valence band profile along an ABC stacked trilayer with co-oriented interfacial polarization. Schematic representation of various parameters used in the self-consistent Poisson-Schrödinger calculation. (b) Variation of bandgap obtained from the frontier orbitals of the self-consistent calculation, as a function of the number of layers for the undoped and electron-doped cases.

9. J. Heyd, G. E. Scuseria, M. Ernzerhof, Erratum: "Hybrid functionals based on a screened Coulomb potential" [J. Chem. Phys. 118, 8207 (2003)]. *J. Chem. Phys.* **124**, 219906 (2006).
10. O. Sinai, L. Kronik, Simulated doping of Si from first principles using pseudoatoms. *Phys. Rev. B* **87**, 235305 (2013).
11. P. Giannozzi *et al.*, QUANTUM ESPRESSO: a modular and open-source software project for quantum simulations of materials. *J. Phys.: Condens. Matter* **21**, 395502 (2009).
12. Y. Wan *et al.*, Room-temperature ferroelectricity in 1T'-ReS$_2$ multilayers. *Phys. Rev. Lett.* **128**, 067601 (2022).
13. Z. Fei *et al.*, Ferroelectric switching of a two-dimensional metal. *Nature* **560**, 336-339 (2018).
14. A. Jindal *et al.*, Coupled ferroelectricity and superconductivity in bilayer Td-MoTe$_2$. *Nature* **613**, 48-52 (2023).
15. C. Hamaguchi, C. Hamaguchi, *Basic semiconductor physics*. (Springer, 2010), vol. 9.
16. H. Peelaers, C. G. Van de Walle, Effects of strain on band structure and effective masses in MoS$_2$. *Phys. Rev. B* **86**, 241401 (2012).
17. A. Kumar, P. K. Ahluwalia, Tunable dielectric response of transition metals dichalcogenides MX$_2$ (M=Mo, W; X=S, Se, Te): Effect of quantum confinement. *Phys. B: Condens. Matter* **407**, 4627-4634 (2012).
28